\documentclass[twocolumn,floatfix,amssymb,aps,superscriptaddress,letterpager]{revtex4-2}
\usepackage[colorlinks=true,urlcolor=blue,citecolor=blue,linkcolor=blue]{hyperref}
\usepackage{graphicx}
\usepackage{color}
\usepackage{epsfig}
\usepackage{amsmath}
\usepackage{amssymb}
\usepackage{mathrsfs}
\usepackage{bm}
\usepackage{subfigure}
\usepackage{float}
\usepackage{url}
\usepackage{enumerate}
\usepackage{braket}
\usepackage{comment}
\usepackage{multirow}
\usepackage{appendix}
\definecolor{violet}{rgb}{0.56,0.0,1.0}
\def\tbib#1{{\bf{#1}}}

\begin{document}

\hsize\textwidth\columnwidth\hsize\csname@twocolumnfalse\endcsname

\title{Long-range spin-orbital order in the spin-orbital SU(2)$\times$SU(2)$\times$U(1) model}

\author{Yang Liu}
\affiliation{School of Physical Science and Technology $\&$ Key Laboratory for Magnetism and
Magnetic Materials of the MoE, Lanzhou University, Lanzhou 730000, China}
\affiliation{Lanzhou Center for Theoretical Physics and Key Laboratory of Theoretical Physics of Gansu Province, Lanzhou University, Lanzhou 730000, China.}

\author{Z. Y. Xie}
\email[]{qingtaoxie@ruc.edu.cn}
\affiliation{Department of Physics,  Renmin University of China,  Beijing 100872,  China}

\author{Hong-Gang Luo}
\affiliation{School of Physical Science and Technology $\&$ Key Laboratory for Magnetism and
Magnetic Materials of the MoE, Lanzhou University, Lanzhou 730000, China}
\affiliation{Lanzhou Center for Theoretical Physics and Key Laboratory of Theoretical Physics of Gansu Province, Lanzhou University, Lanzhou 730000, China.}
\affiliation{Beijing Computational Science Research Center, Beijing 100084, China}

\author{Jize Zhao}
\email[]{zhaojz@lzu.edu.cn}
\affiliation{School of Physical Science and Technology $\&$ Key Laboratory for Magnetism and
Magnetic Materials of the MoE, Lanzhou University, Lanzhou 730000, China}
\affiliation{Lanzhou Center for Theoretical Physics and Key Laboratory of Theoretical Physics of Gansu Province, Lanzhou University, Lanzhou 730000, China.}

\begin{abstract}
By using the tensor-network state algorithm, we study a spin-orbital model with SU(2)$\times$SU(2)$\times$U(1) 
symmetry on the triangular lattice. This model was proposed to describe some triangular $d^1$ materials and was argued to 
host a spin-orbital liquid ground state. In our work the trial wavefunction of its ground state is approximated by an 
infinite projected entangled simplex state and optimized by the imaginary-time evolution. 
Contrary to the previous conjecture, we find that the two SU(2) symmetries are broken, resulting in a stripe spin-orbital order with 
the same magnitude $m=0.085(10)$. This value is about half of that in the spin-1/2 triangular Heisenberg antiferromagnet. 
Our result demonstrates that although the long-sought spin-orbital liquid is absent in this model the spin-orbital order is significantly reduced due to 
the enhanced quantum fluctuation. This suggests that high-symmetry spin-orbital models are promising in searching for
exotic states of matter in condensed-matter physics.
\end{abstract}

\maketitle
\textit{Introduction.} 
Symmetry may be one of the fundamental concepts involved in physics. Of all these symmetries SU(2) is ubiquitous 
in condensed-matter physics because the spin operators are its generators. 
On the other hand, the success of larger symmetry groups such as SU(3) in particle physics 
raised a natural question whether high-symmetry groups are relevant for condensed-matter physics.
It was proposed that high symmetries can be realized in spin-orbit-coupled compounds~\cite{Khomskii-SPU1982, ZFC-PRL1998}.
This has inspired a variety of theoretical studies on spin-orbit-coupled insulators \cite{Nersesyan-PRL1999, Affleck-PRB2000, Ueda-JPSJ2000, Xiang-PRB2008, Sirker-PRB2011, Mila-PRX2012, Vishwanath-PRB2009, Li-PRB2017, Chen-PRB2022}. 
At the meanwhile, the experimental development of measuring the orbital degrees of freedom in transition-metal 
oxides \cite{Tokura-PRL1998, Singh-Nat2012} has greatly boosted the 
investigation on orbital physics in materials with strong spin-orbit coupling and crystal-field splitting. 
A minimal high-symmetry model involving both the spin and orbital degrees of freedom may be the SU(4) symmetric Kugel-Khomskii model~\cite{Khomskii-SPU1982, ZFC-PRL1998, Mila-PRX2012, Jackeli-PRL2018, Jackeli-PRB2021, Jian-PRL2020, Zhou-SB2022}, where orbital degrees of freedom are represented as pseudospin and coupled with the spin degree of freedom on each bond.
It is argued that such a model is relevant to the observed spin liquid states~\cite{Savary-RPP2016, Zhou-RMP2017, Senthil-Sci2020} in $\rm{LiNiO_2}$~\cite{ZFC-PRL1998},  $\rm{Ba_3CuSb_2O_9}$ \cite{Mila-PRX2012}, 
and $\alpha$-$\rm{ZrCl_3}$~\cite{Jackeli-PRL2018}. 

On the other hand, Yamada, Oshikawa, and Jackeli~(YOJ) recently proposed~\cite{Jackeli-PRB2021} another spin-orbital model to 
describe some $d^1$ materials~\cite{Lee-PNAS2017, LiSY-PRB2017, Shigeta-RSCA2018} on the triangular lattice. 
In addition to the geometry frustration, there is a spin-orbital frustration in this model, and it is argued~\cite{Jackeli-PRB2021} that such a model may host a spin-orbital liquid ground state.
However, there are no systematic studies so far, which motivates us to investigate this model. 

\begin{figure*}[htbp]
	\centering
	\includegraphics[width=0.98\textwidth, clip]{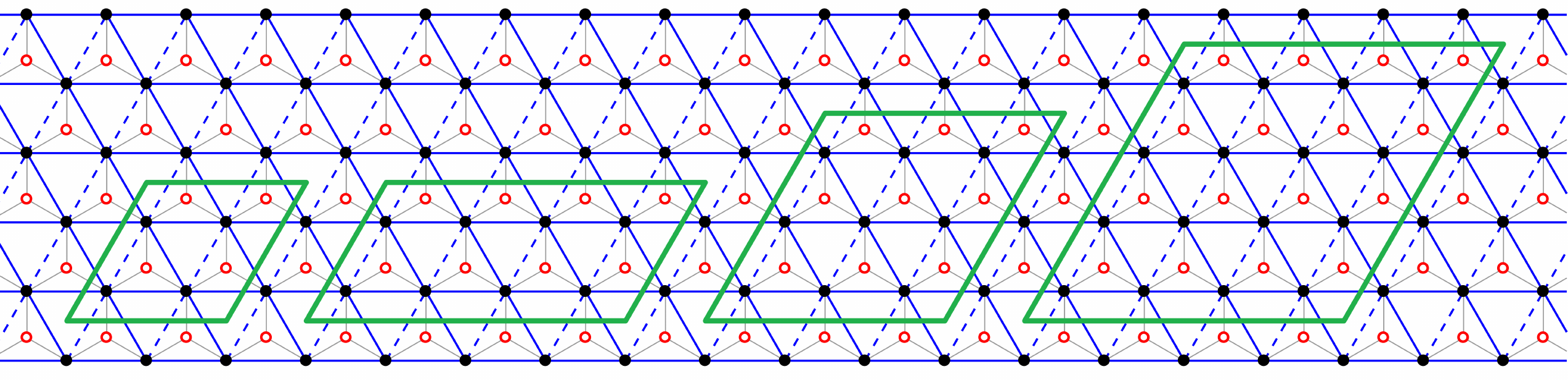}
	\caption{Schematic diagram of the triangular lattice and the PESS wavefunction ansatz. The black filled circles ($\bullet$) represent the lattice sites. 
		The blue solid and dashed lines connecting two nearest-neighbor lattice sites are the bonds of the lattice, 
		indicating two different interaction terms corresponding to Eqs.~(\ref{SU4}) and (\ref{quasi-SU4}), respectively.
		The red open circles ($\color{red}{\bm{\circ}}$) sitting at the center of upward triangles represent the rank-3 
		simplex tensors $S$ in the PESS wavefunction, and at each lattice site there is a projection tensor $A$. 
		The physical indices $\sigma$ and $\sigma'$ perpendicular to the plane are not shown here.
		The four rhombuses in green mark the $2\times 2$, $4\times 2$, $3\times 3$, and $4\times 4$ periodic clusters used in the trial wave functions.}
	\label{3-PESS}
\end{figure*}

In the YOJ model, there are two kinds of terms in the Hamiltonian, as plotted in Fig.~\ref{3-PESS}. 
The lattice sites are represented by filled circles. The nearest-neighbor sites are connected by blue solid or dashed lines, which represent 
the two interaction terms, respectively. The solid lines represent the term 
\begin{equation}
        H_{ij}=\left( \boldsymbol S_{i} \cdot \boldsymbol S_{j} + \frac{1}{4}\right)\left(\boldsymbol T_{i} \cdot \boldsymbol T_{j} + \frac{1}{4}\right),
        \label{SU4}
\end{equation}
where $\boldsymbol{S_i}$ and $\boldsymbol{T_i}$ are both pseudospin-1/2 operators defined on lattice site $i$, 
corresponding to effective spin and orbital degrees of freedom, respectively. 
The dashed lines are the term given by  
\begin{equation}
        H'_{ij} = \left( \boldsymbol S_{i} \cdot \boldsymbol S_{j} + \frac{1}{4}\right)\left(T^{z}_{i}T^{z}_{j}-T^{x}_{i}T^{x}_{j}-T^{y}_{i}T^{y}_{j} + \frac{1}{4}\right).
        \\\label{quasi-SU4}
\end{equation}
The Hamiltonian $\mathcal{H}$ of the YOJ model is then the summation of all these terms. 

\textit{Symmetry and method.}  
Although Eq.~(\ref{quasi-SU4}) breaks the orbital SU(2) symmetry to U(1), it turns out that this model has
a SU(2)$\times$SU(2)$\times$U(1) symmetry. To see this, one can construct~\cite{Zhang-PRB1999} the generators $\boldsymbol{X}_i$ and $\boldsymbol{Y}_i$ as follows,
\begin{equation}
	X^\alpha_i \equiv S^\alpha_i\left(\frac{1}{2} + T^z_i\right),  \quad Y^\alpha_i \equiv S^\alpha_i\left(\frac{1}{2} - T^z_i\right)
	\label{EffXY}
\end{equation}
where $\alpha = x,y,z$. It can be easily verified that  
\begin{align}
	[X^\alpha_i, X^\beta_j] &= i\varepsilon_{\alpha\beta\gamma}X^{\gamma}_i\delta_{ij}, \nonumber \\
	[Y^\alpha_i, Y^\beta_j] &= i\varepsilon_{\alpha\beta\gamma}Y^{\gamma}_i\delta_{ij}, \nonumber \\
	[X^\alpha_i, Y^\beta_j] &= 0, \nonumber \\
	[T^z_i, X^\alpha_j]     &= 0, \nonumber \\
	[T^z_i, Y^\beta_j]      &= 0
\end{align} 
are satisfied. Moreover, the following seven operators are conserved quantities with respect to the Hamiltonian,
\begin{equation}
	X^\alpha \equiv \sum_{i}X^\alpha_i, \quad Y^\alpha \equiv \sum_{i}Y^\alpha_i, \quad T^z \equiv \sum_{i}T^z_i.
\end{equation}
Knowing the commutation relation of these operators, we have found seven generators of a Lie group which can be written as SU(2)$\times$SU(2)$\times$U(1), 
and we can use these generators to define the corresponding order parameters as usually done for spin operators \cite{LQ-PRB2022}.  

To study the ground-state properties of this model, we employ the tensor-network method \cite{Nishino2001, PEPS2004, SimBook2018, OrusReview}, which is a class of numerical methods based on the tensor-network state representation of the targeted quantum state and the network contraction techniques arising from the idea of renormalization group.
It is free of the sign problem, can study the thermodynamic limit directly under the help of translational invariance, and has been successfully applied to study strongly-correlated electron systems \cite{tJ2014, Simons-XXS}, frustrated spin systems \cite{Corboz-PRL2013, WL-PRB2016, Liao-PRL2017, LQ-PRB2022}, statistical models \cite{HOTRG-PRB2012, CW-PRB2014, ZP-PRL2021}, topological order \cite{Wen-PRB2009, Vidal-PRB2009, Eisert-PRB2017, ZGM-PRL2020, Frank-arXiv2017}, quantum field theory \cite{CMPS2010, LQCD2013, CTNS2019}, machine learning \cite{ML2018, ML2020}, and even quantum circuit simulation \cite{Yannick-RMP2022}, etc. In this work, we use the infinite projected entangled simplex state (PESS) ansatz \cite{PESS2014, Schuch-PRB2012} to represent the trial wavefunction, and employ the corner transfer-matrix renormalization group (CTMRG) method \cite{CTMRG1996, CTMRG2009, tJ2014} combined with the nested tensor network technique \cite{NTN-PRB2017} to estimate the physical quantities, such as the ground-state energy and order parameters, etc. 
Unexpectedly, we find that in the ground state two SU(2) symmetries are broken. As we have shown, 
the generators of two SU(2) groups include both the spin and orbital operators, 
which suggests that the corresponding orders are different from the conventional magnetic order~\cite{LQ-PRB2022}. 
Hereafter, following the literature~\cite{Jackeli-PRL2018, Jackeli-PRB2021} we call it a spin-orbital order. 

More specifically, the PESS employed in this work is a generalization of the projected entangled pair state ansatz \cite{PEPS2004}, 
and has been successfully applied to the highly frustrated antiferromagnetic Heisenberg model on Kagome lattice \cite{PESS2014, Liao-PRL2017, YJK-PRB2018} 
and triangular lattice \cite{LQ-PRB2022}.  Similar to that in Ref.~\cite{LQ-PRB2022}, in this work, the PESS wavefunction is represented as follows,
\begin{align}
	|\Psi\rangle =\sum_{\{\sigma,\sigma'\}}\mathrm{Tr}&\left(...S^{(\mu\nu)}_{i_{\mu\nu}j_{\mu\nu}k_{\mu\nu}}
	A^{(\lambda\omega)}_{i_{\lambda\omega}j_{\lambda\omega}k_{\lambda\omega}}
	[\sigma_{\lambda\omega}\sigma'_{\lambda\omega}]...\right) \nonumber \\
	&|...\sigma_{\lambda\omega}...\rangle|...\sigma'_{\lambda\omega}...\rangle
	\label{Eq:3-pess}
\end{align}
as illustrated in Fig.~\ref{3-PESS}. Here $(\mu,\nu)$ denotes the coordinates of the upward triangles, at the center of which a rank-3 simplex tensor $S$ is introduced to characterize the entanglement in that triangle. $(\lambda,\omega)$ denote the coordinates of lattice sites, where a rank-5 projection tensor $A$ is defined, with three virtual indices labeled as $i, j, k$ and two physical indices labeled as $\sigma$ (spin) and $\sigma'$ (orbital). Every two virtual indices associated with the same bond take the same values. ``$\mathrm{Tr}$'' is over all the repeated virtual indices and $\sum$ is over all the physical indices.

The bond dimension $D$, which is the highest value that the virtual indices can take, is an important parameter in tensor network states. Generally, the larger $D$ is, the more accurate the obtained representation is, but the heavier the computational cost is at the meanwhile. Therefore, in practical calculations, one has to make a good balance between accuracy and cost. In this work, using the efficient nested tensor network technique \cite{NTN-PRB2017}, we have pushed $D$ to 18 and obtained reliable features of the ground state.

The ground-state wavefunction is obtained by imaginary-time evolution. In order to make the calculation more efficient, the wavefunction is updated 
by the simple update algorithm \cite{SU1D2007, SU2D2008}. Though for a given $D$, simple update might be less accurate than 
the full update \cite{FU2014} or direct variational calculation \cite{SRGAD-DTNS}, it can produce wavefunction with a much larger $D$ and the numerical 
accuracy can thus be remedied properly, as exemplified in Ref.~\cite{Liao-PRL2017}. Moreover, to avoid bias and reduce the Trotter error, 
we start from a wavefunction randomly generated on complex field, and gradually reduce the Trotter step $\tau$ from $0.5$ to $10^{-3}$, 
which turns out to be sufficiently small to estimate the physical quantities.

After mapping the obtained infinite PESS wavefunction to a square network \cite{Note1}, we measure the physical observables through the CTMRG method developed for an arbitrary unit cell on the square lattice \cite{tJ2014}. The central idea of this method is to represent the surrounding environment of a local tensor in terms of corner matrices and edge tensors approximately. A key parameter therein is the bond dimension $\chi$ of the environment tensors, 
which controls both the accuracy and cost during the measurement. In principle, our results should also depend on $\chi$. 
We will show later such dependence is very weak and thus it can be neglected when the $\chi$ is large enough, i.e., $\chi\ge{D^2}$. 
In our calculation, the maximal $\chi$ is no less than $D^2$ to ensure a reliable result.  

\begin{figure}[!ht]
    \includegraphics[width=0.99\columnwidth, clip]{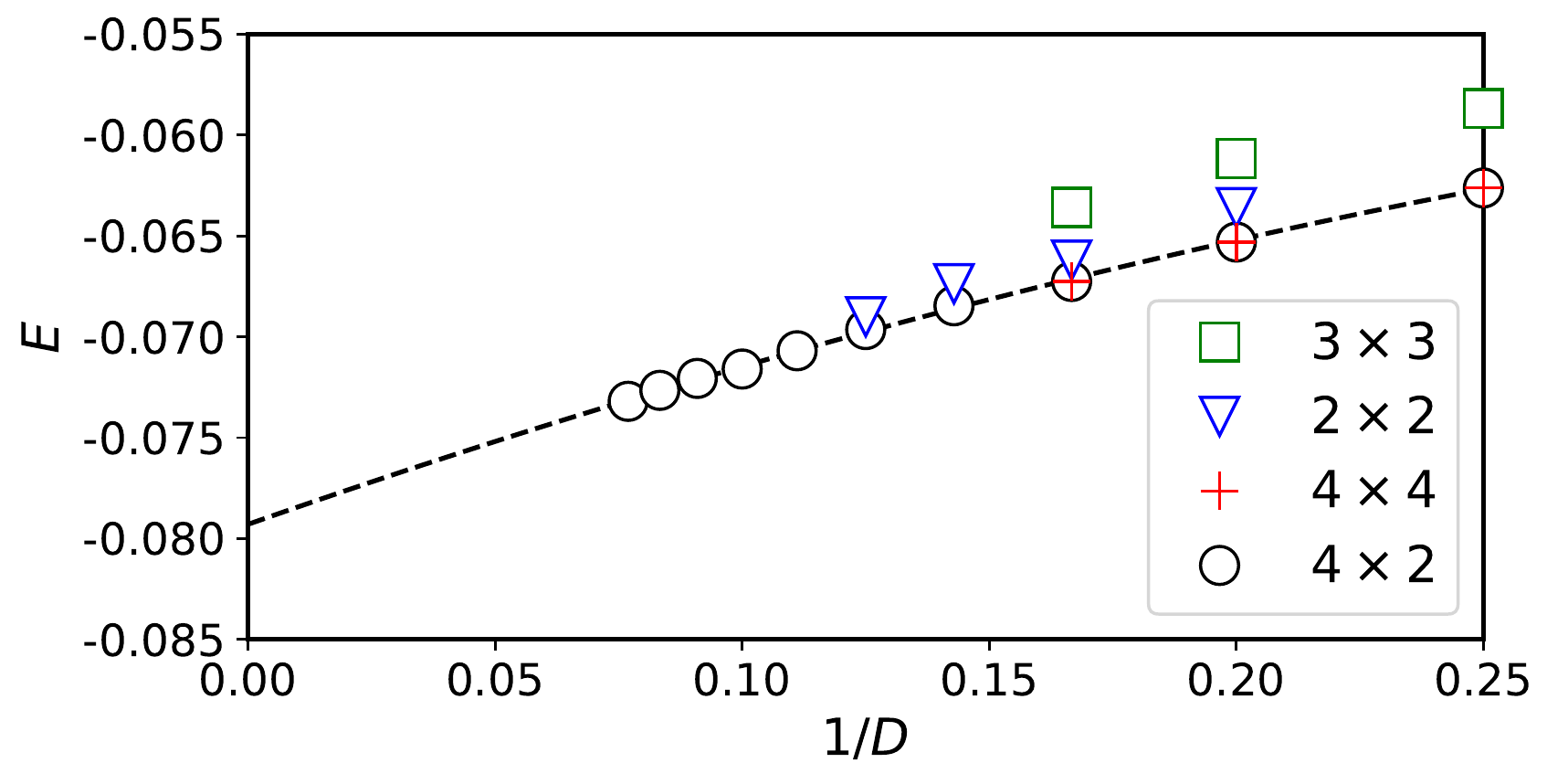}\\
    \caption{The energy per bond $E$ obtained from the wavefunction ansatz with different clusters are shown as a function of $1/D$.}
    \label{EnergyCmp}
\end{figure}

\textit{Ground-state energy.} 
In the tensor-network simulations, the cluster size should be compatible with the unit cell if the ground state has a long-range order. 
However, since the ground state is unknown a priori, we need to try different cluster sizes to 
determine the unit cell. A correct cluster size should have the lowest energy in correspondence to the ground state. 
For this purpose, we compare the energy obtained from the PESS ansatz with various clusters. 
We have checked four different clusters compatible with the Hamiltonian symmetry, i.e., $2\times2$, $3\times3$, $4\times2$, $4\times4$, 
which are all illustrated in Fig.~\ref{3-PESS}. The energy $E$ obtained from the ansatz with these four clusters is shown in Fig.~\ref{EnergyCmp} 
for comparison. It shows that the energy obtained on the $2\times2$ and $3\times3$ clusters is obviously higher than that 
on the $4\times2$ and $4\times4$ clusters. Moreover, the energy is the same on the latter two clusters, suggesting that the 
$4\times2$ is the minimal cluster compatible with the ground state. Therefore, in all the rest calculations, we focus on the 
ansatz on the $4\times2$ cluster.

In Fig.~\ref{EnergyCmp}, we extrapolate the energy per bond $E$ as a function of $1/D$ to the large-$D$ limit by the formula 
\begin{eqnarray}
	E(1/D)=E_g+a_0(1/D)+a_1(1/D)^2,
\end{eqnarray}
where $E_g$, $a_0$ and $a_1$ are the fitting parameters. We finally end with the ground-state energy per bond $E_g = -0.0793(5)$. 
It may serve as a benchmark for future works.

\textit{Spin-orbital order.}
It is argued that the quantum fluctuation is strong~\cite{Zaanen-PRL1997, Wu-PRL2005, Jackeli-PRL2018} 
in high-symmetry models and thus spin-orbital liquid is favored therein. 
This argument is supported by several works~\cite{Mila-PRX2012, Jian-PRL2020, Zhou-SB2022} on the SU(4) spin-orbital model on various lattices. In this context, 
the main concern on this model is whether its symmetries are broken or not, in particular, whether the SU(2)$\times$SU(2)$\times$U(1) symmetry is broken or not. 
To check this, we calculate the corresponding local order parameters, which are defined as the expectation values of the generators in the ground state. 
The magnitude of the spin-orbital order at site $i$ is defined by $m_i^{X}=\sqrt{\sum_\alpha{\langle X_i^\alpha\rangle}^2}$ 
and $m_i^{Y}=\sqrt{\sum_\alpha{\langle Y_i^\alpha\rangle}^2}$ corresponding to two SU(2) groups. Their average in one unit cell is then defined 
by $m^{X/Y}=\sum_i{m_i^{X/Y}}/N$ with $N=4\times{2}=8$. 

\begin{figure}[!ht]
        \includegraphics[width=\columnwidth, clip]{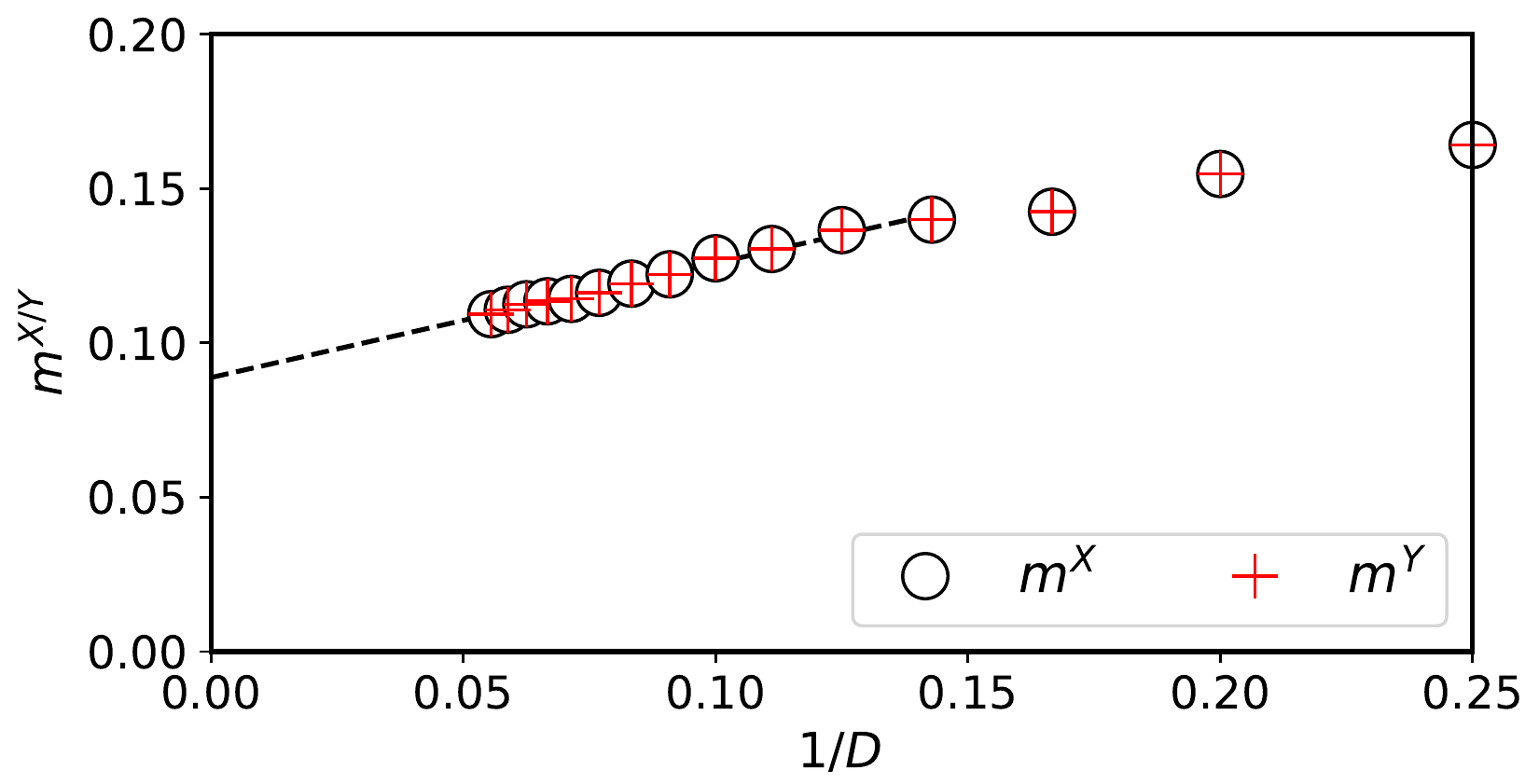}\\
        \caption{The magnitude of the spin-orbit order obtained with $D$ ranging from 4 to 18. Various polynomial functions are used to fit the curve. The dashed line denotes a linear fitting. It is estimated that $m^{X/Y}=0.085(10)$ in the large-$D$ limit.}
        \label{Mag}
\end{figure}
In Fig.~\ref{Mag}, we show $m^{X/Y}$ as a function of $1/D$ with $D$ ranging from 4 to 18. One may notice that $m^X$ and $m^Y$ coincide very well.
This can be understood as follows. Let us check the two SU(2) groups. Although their generators $X_i^\alpha$ and $Y_i^\beta$ commute, we can see that 
$X_i^\alpha\leftrightarrow{Y_i^\alpha}$ under the transformation $T_i^z\rightarrow{-T_i^z}$.  
In addition, $m^{X/Y}$ decreases monotonically as $D$ increases. We try to fit such a curve by various polynomial functions of $1/D$,  
which gives $m^{X/Y}=0.085(10)$ in the large-$D$ limit. Obviously, the finite magnitudes tell us that two SU(2) symmetries 
are spontaneously broken. Moreover, the magnitudes are about half of that of the spin-1/2 anfiferromagnetic 
Heisenberg model~(see, for example, Table I in Ref.~\cite{LQ-PRB2022}).
This is consistent with the argument that quantum fluctuation is enhanced~\cite{Zaanen-PRL1997, Wu-PRL2005, Jackeli-PRL2018} in high-symmetry models. 
In addition, $|\langle T^z\rangle|=\sum_i{|\langle T_i^z \rangle|}/N$ is always zero within our error bar, suggesting that 
the U(1) symmetry is not broken.   

\begin{figure}[htbp]
			\includegraphics[width=0.36\textwidth, clip]{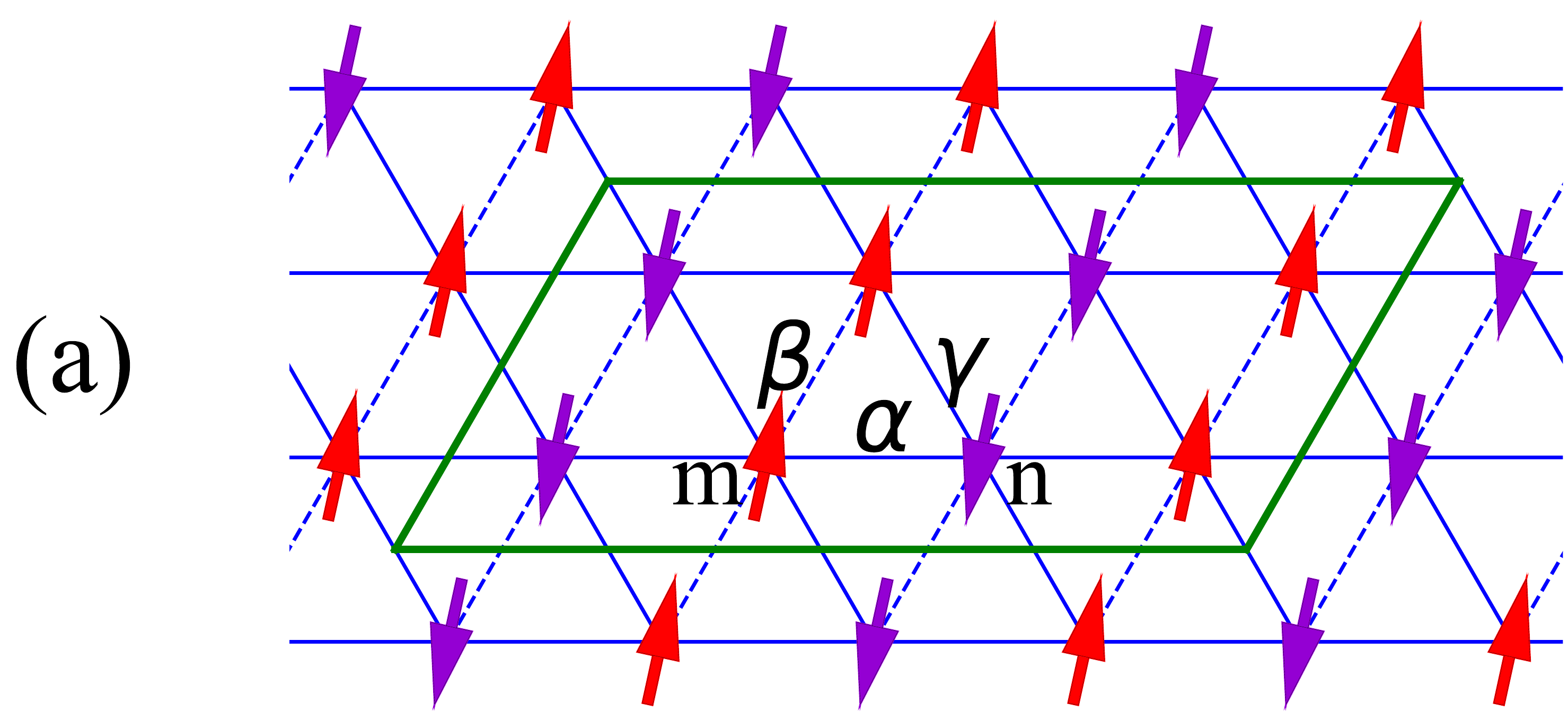}
			\includegraphics[width=0.36\textwidth, clip]{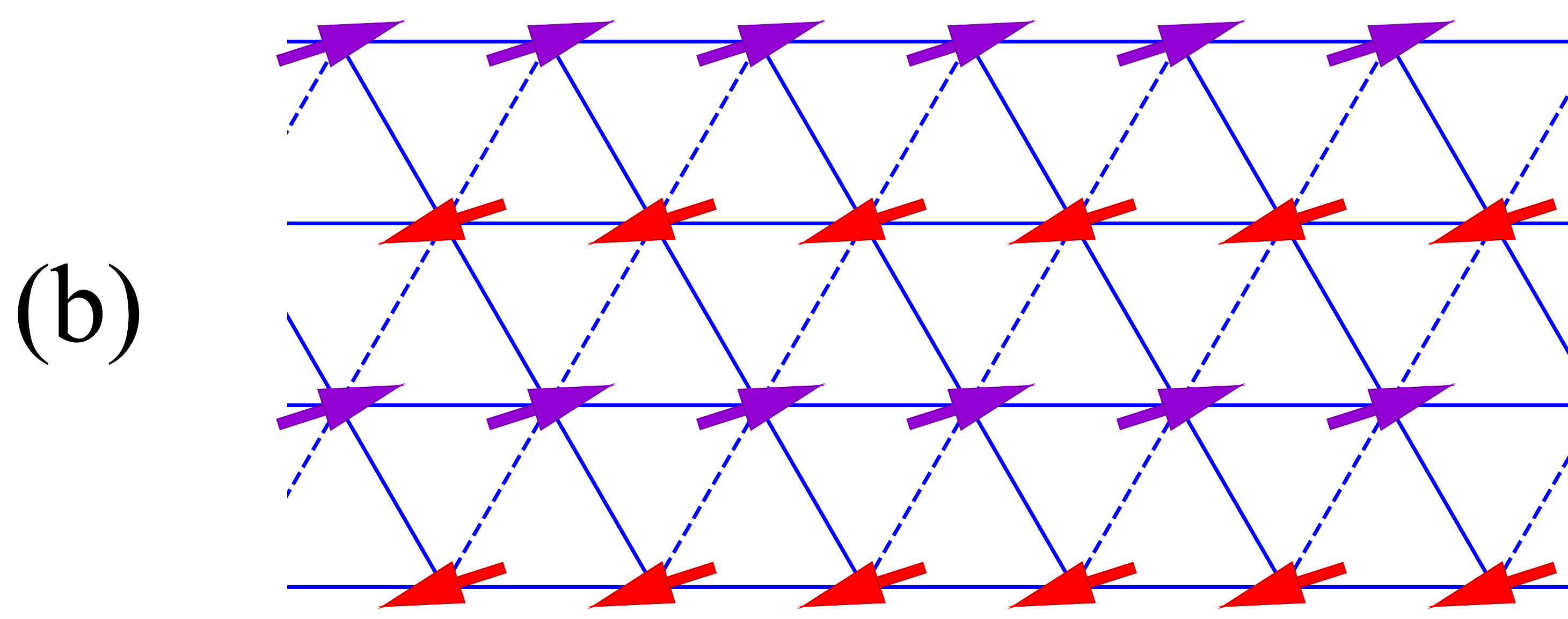}
			\includegraphics[width=0.36\textwidth, clip]{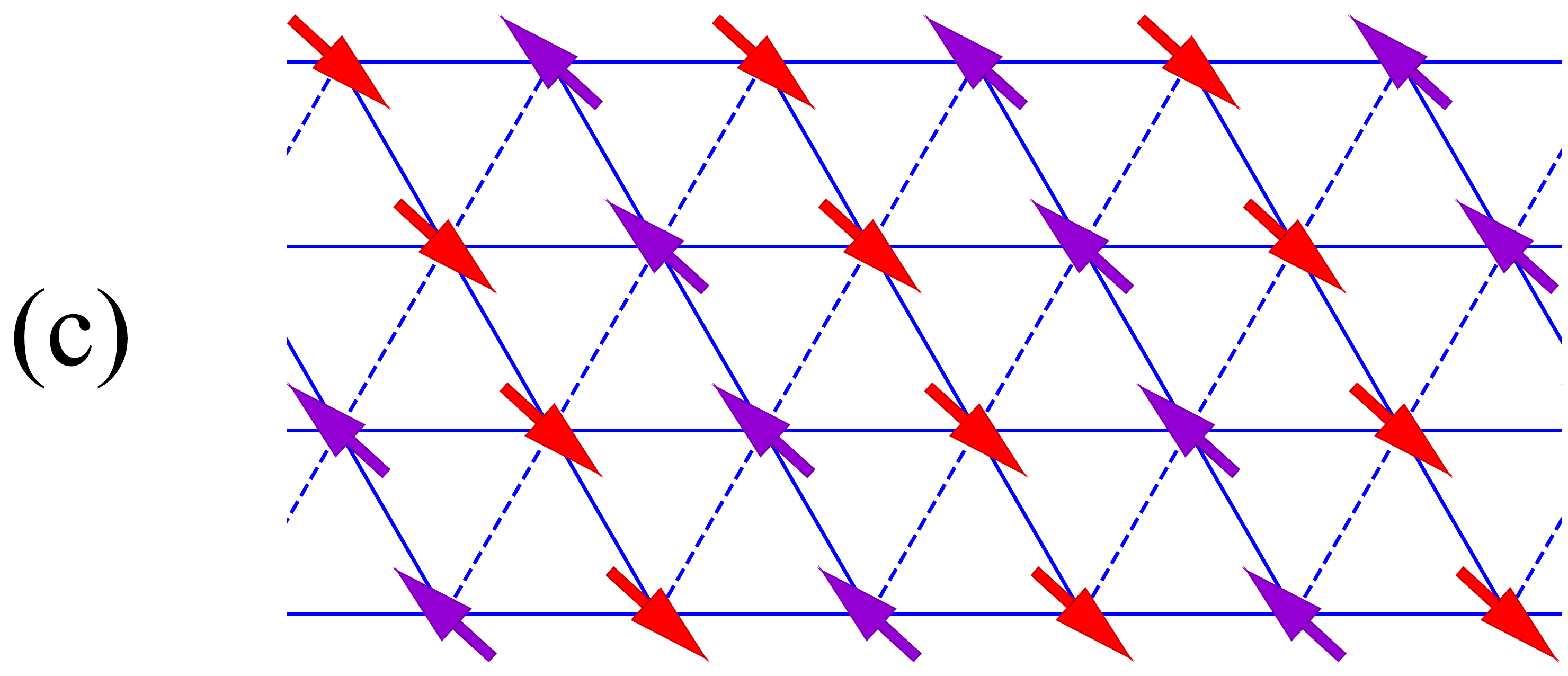}
	\caption{Degenerate stripe spin-orbital configurations for the ground state. The data are obtained with $D=18$ and $\chi=D^2$.}
	\label{MagStr}
\end{figure}

Now we have demonstrated that there is a stable long-range spin-orbital order in the YOJ model. 
Next, we will show the pattern of such an order. In Fig.~\ref{MagStr} 
the spin-orbital order corresponding to the three components of $\boldsymbol{X}_i$ 
is plotted as a vector. It shows clearly that the ground state exhibits a stripe long-range order. 
From Fig.~\ref{MagStr}(a) we can see that along the dashed bonds, 
where the interaction is given in Eq.~(\ref{quasi-SU4}), the spins are ferromagnetic, and along the two solid bonds the spins are antiferromagnetic.
However, we want to point out that the spins can be ferromagnetic along one of two solid bonds, i.e., 
(b) and (c) are also the ground-state configurations. 
This can be understood as follows. First, let us see the equilateral triangle in Fig.~\ref{MagStr} with its three sides $\alpha$, $\beta$, and $\gamma$. 
The lines $\alpha$ and $\beta$ join at the site, say, $m$. Lines $\alpha$ and $\gamma$ join at site $n$. After the transformation 
$T_m^x\rightarrow -T_m^x, T_m^y\rightarrow -T_m^y$, $T_n^x\rightarrow -T_n^x, T_n^y\rightarrow -T_n^y$, the side $\beta$ becomes solid 
and side $\gamma$ becomes dashed.
By a similar operation, all the bonds parallel to side $\beta$ can become solid, and all bonds parallel to side $\gamma$ can be dashed. 
Keep in mind that our order parameters do not depend on $T_i^x$ and $T_i^y$, which means their expectation values do not change. We then rotate the lattice 
counterclockwise by $2\pi/3$ around an axis perpendicular to the plane at any lattice site, and we obtain the figure in panel (b). 
Similarly, we can obtain the configuration in panel (c).

\textit{Convergence analysis.}
In order to obtain  reliable data, we not only pushed the bond dimension $D$ as large as possible, 
but also checked the convergence of the expectation values with respect to the environment tensor dimension $\chi$ for {each $D$ 
as in Refs.~\cite{NTN-PRB2017, Liao-PRL2017, LQ-PRB2022}. For example, in Fig.~\ref{ChiScale} we plot $m^{X}$ 
as a function of $\chi$ for several $D$s. It shows that the data converge quickly as $\chi$ increases, which demonstrates that $\chi\sim D^2$ 
is sufficiently large to produce a reliable result for all $D$s up to 18. Therefore, with $\chi\ge{D^2}$ our results should be reliable.

\begin{figure}[ht]
	\includegraphics[width=0.5\textwidth, clip]{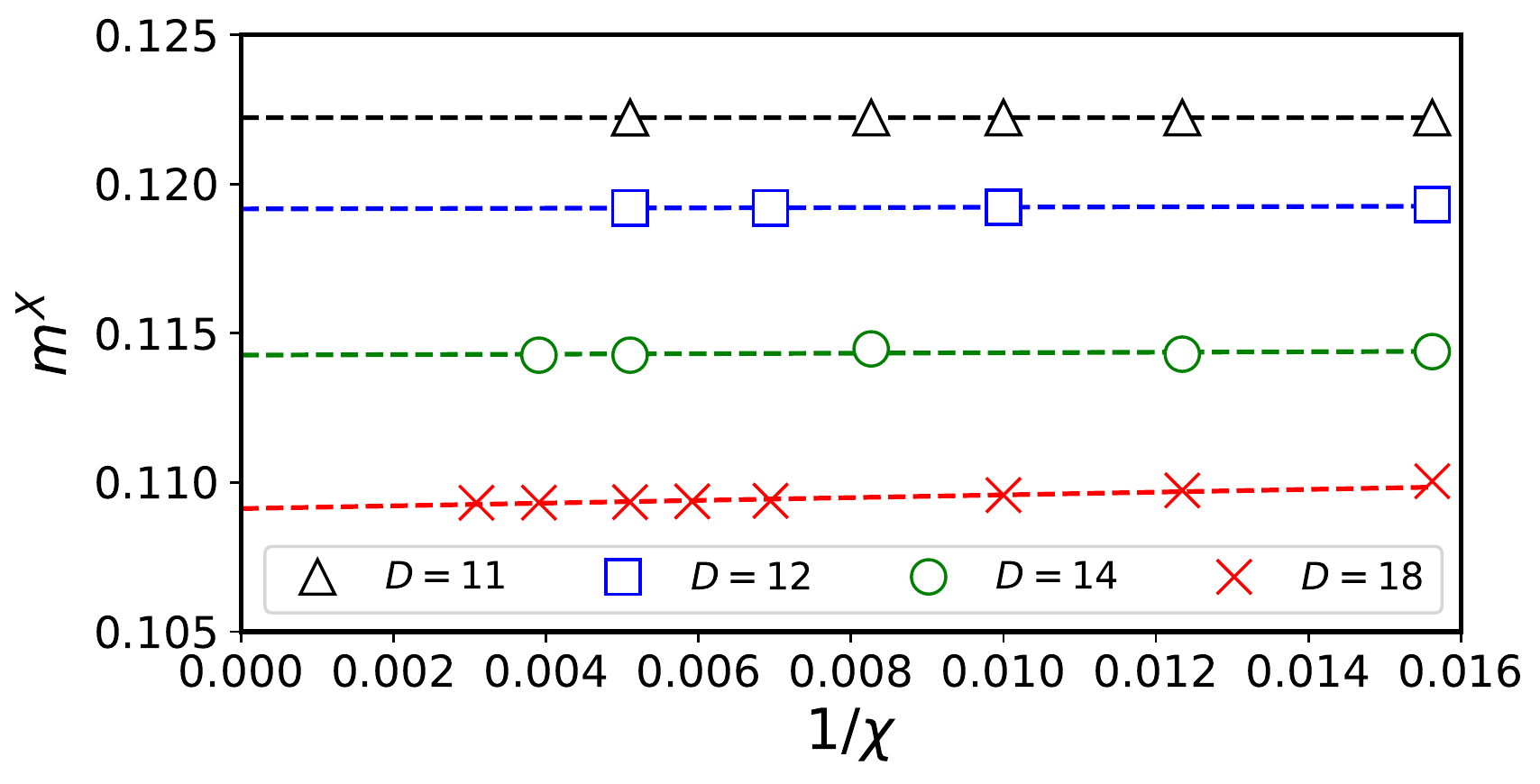}
	\caption{$m^X$ is shown as a function of $1/\chi$ for $D = 11, 12, 14$ and $18$. }
	\label{ChiScale}
\end{figure}

\textit{Conclusion}.
In summary, by using tensor-network algorithms with PESS wavefunction ansatz~\cite{PESS2014, Schuch-PRB2012}, we have studied the ground state 
of the spin-orbit-coupled YOJ model~\cite{Jackeli-PRB2021} on the triangular lattice, which possesses a SU(2)$\times$SU(2)$\times$U(1) symmetry.
The trial wavefunction was optimized by the imaginary-time evolution method,
and the expectation values were estimated by the multi-sublattice CTMRG algorithm in combination with the nested tensor-network technique. 
We found that the two SU(2) symmetries are broken, leading to long-range spin-orbital orders with a stripe pattern. 
The origin of these orders is different from the conventional magnetic order. 
A careful finite bond-dimension scaling analysis gives the magnitudes of the spin-orbital orders $m^X=m^Y=0.085(10)$. The reason for $m^X=m^Y$ is also discussed. 
Our resuls impose a strong constraint on the microscopic Hamiltonian in searching for quantum spin-orbital liquid in real materials on the triangular lattice. 

We are grateful to Hong-Hao Tu and Shaojin Qin for helpful discussions and  Ken Chen for plotting Fig. 1.
This work is supported by the National key R$\&$D Program of China (Grants No. 2022YFA1402704, No. 2017YFA0302900 and No. 2016YFA0300503), by the National Natural Science Foundation of China (Grants Nos. 12274187, 12274458, 12047501, 11874188, 11834005, 11774420), and by the Research Funds of Renmin University of China (Grants No. 20XNLG19).

\end{document}